\documentclass[12pt]{article} \textheight=24 true cm
\usepackage{graphicx}
\usepackage{amsmath}
\usepackage{amssymb}
\usepackage{color}
\usepackage[colorlinks]{hyperref}

\begin{document}

 \begin{center}
{\Large\bf Inhomogeneities in  dusty universe - a possible
alternative to dark energy?
}\\[8 mm]
S. Chatterjee\footnote{Relativity and Cosmology Research Centre,
Jadavpur University, Kolkata - 700032, India, and also at IGNOU,
New Alipore College, Kolkata  700053, e-mail : chat\_
sujit1@yahoo.com }\\[6mm]
\end{center}

\begin{abstract}
There have been of late renewed debates on the role of
inhomogeneities to explain the observed late acceleration of the
universe. We have  looked into the problem analytically with the
help of the well known spherically symmetric but inhomogeneous
Lemaitre-Tolman-Bondi(LTB) model generalised to higher dimensions.
It is observed that in contrast to the claim made by Kolb et al
the presence of inhomogeneities as well as extra dimensions can
not reverse the signature of the deceleration parameter if the
matter field obeys the energy conditions. The well known
Raychaudhuri equation also points to the same result. Without
solving the field equations explicitly it can, however, be shown
that although the total deceleration is positive everywhere
nevertheless it does not exclude the possibility of having radial
acceleration, even in the pure dust universe, if the angular scale
factor is decelerating fast enough and vice versa. Moreover it is
found that introduction of extra dimensions can not reverse the
scenario. To the contrary it actually helps the decelerating
process.
\end{abstract}

 ~~~Keywords : accelerating universe ; inhomogeneity; higher dimensions

~~ PACS: 04.20, 04.50 +h

\section{INTRODUCTION}

Following three discoveries in the last century our concept of
cosmological evolution has undergone a sea change - in contrast to
Einstein's idea of a static universe Hubble and Slipher(1927)
established that it is in fact expanding. Further the CMBR spectra
as well as primordial nucleosynthesis studies in the sixties show
that the observable universe was in an extremely hot dense state
in the past and has since been expanding for the last 13.5 Gyr. as
dictated by Einstein's theory. Finally from the high redshift
supernovae data in the last decade  \cite{riess} we know that when
interpreted within the framework of the standard FRW type of
universe(homogeneous and isotropic) we are left with the only
alternative that the universe is now going through an accelerated
expansion with baryonic matter contributing only five percent of
the total budget. Later data from CMBR studies\cite{spergel}
further corroborate this conclusion which has led a vast chunk of
cosmology community (\cite{xx} and references therein) to embark
on a quest to explain the cause of the acceleration. The teething
problem now confronting researchers is the identification of the
mechanism that triggered the late inflation. Workers in this field
are broadly divided into two groups- either modification of the
original general theory of relativity or introduction of any
mysterious fluid in the form of an evolving cosmological constant
or a quintessential type of scalar field. So far the main emphasis
for explaining the recent acceleration rests on the assumption of
a homogeneous FRW type of model. But measurements of average
matter density from the different cosmic probes on supernovae
($\Omega_{M}\sim 0$), galaxy distributions ($\Omega_{M}\sim 0.3 $)
and CMBR ($\Omega_{M}\sim 1$) point to a highly confusing picture
of the universe such that at least one of the two assumptions must
be wrong. As is well known theorists attempt to evade the problem
by introducing a cosmological constant(vacuum energy) such that
values now become for best fits as supernovae ($\Omega_{M}\sim
0.3$ and $\Omega_{\Lambda}\sim 0.7$), galaxy distributions
($\Omega_{M}\sim 0.3$) and CMBR ($\Omega_{M} +
\Omega_{\Lambda}\sim  1)$. But the popular explanation with the
help of a cosmological constant is beset with serious theoretical
problems because absence of acceleration at redshifts $z\geq1$
implies that the required value of the cosmological constant is
approximately 120 orders of magnitude smaller than its natural
value in terms of Planck scale \cite{copeland}. As for the
alternative quintessential field  we do not in fact have a theory
that would explain, not to mention predict, the existence of a
scalar field fitting the bill without violating the realistic
energy conditions. Moreover we can not generate this type of a
scalar field from any basic principles of physics.  So there has
been a resurgence of interests among relativists, field theorists,
astrophysicists and people doing astroparticle physics both at
theoretical and experimental levels to address the problems
emanating from the recent extra galactic observations without
involving a mysterious form of scalar field by hand but looking
for alternative approaches  based on sound physical principles.
Alternatives include, among others, higher curvature theory,
axionic field and also Brans- Dicke field. Some people attempted
to look into the problem from a purely geometric point of view -
an approach more in line with Einstein's spirits. For example,
Wanas \cite{wanas} introduced torsion while Neupane \cite{nepune}
modifies the spacetime with a  warped factor in 5D spacetime in a
brane like cosmology and finally addition of extra spatial
dimensions in physics \cite{sc}. But the problem with Wanas' model
is that the  the geometry is no longer Riemannian. On the
otherhand the fact that  the warped spacetime always generates
acceleration is fairly wellknown and it follows from the wellknown
Raychaudhuri equation also. Moreover all the conceptual problems
relating to brane models are present in Neupane's model.\\ In this
context one important thing should not escape our attention. One
intriguing fact in the framework of the standard FRW model is that
the accelerating phase coincides with the period in which
inhomogeneities in the matter distribution at length scales $<¡«
10$ Mpc become significant so that the Universe can no longer be
approximated as homogeneous  at these scales. One should note that
homogeneity and isotropy of the geometry are not essential
ingredients to establish a number of relevant results in
relativistic cosmology. One need not be sacrosanct about these
concepts so far as the relativistic cosmology is concerned. For
instance, from the early sixties to the early seventies
(\cite{lifsitz} and references therein), a research program on the
singularity properties of general cosmological solutions has been
conducted without relying on the isotropy and on the homogeneity
of the geometry. The theme of the present article is somehow
opposite to the one analysed in \cite{lifsitz} where the emphasis
was on the role of the inhomogeneities (and anisotropies) in the
proximity of a cosmological singularity. A link between
inhomogeneities and cosmological acceleration has been pursued in
various studies in recent past, although the journey is not free
from serious controversies creeping up time to time. There have
been arguments, based on perturbative estimates, that the
backreaction of super horizon inhomogeneities on the cosmological
expansion is significant and could cause acceleration \cite{kolb}
when observed from the centre of perturbation. This idea is later
supplemented by Wiltshire \cite{wil} and also Carter et al
\cite{car} where the observed universe is assumed to be  an
underdense bubble in an Einstein-de Sitter universe and it was
shown that from observational point of view their results become
very similar to the predictions of $\Lambda CDM$ model. However,
the validity of the perurbative ansatz is questionable in that the
claimed acceleration is later shown to be due to the result of
extrapolation of a specific solution to a regime where both the
perturbative expansion breaks down and and the constraints are
violated \cite{giovannini}. Again we know \cite{book} that in a
matter dominated nonrotating model where particles interacting
with one another move along geode4sic lines it is always possible
to define a coordinate system which is at once
synchronous($g_{00}=1$) and comoving. With this input Hirata and
Seljak \cite{hirata} conclusively showed from Ray Chaudhuri
equation that in a perfect fluid cosmological model that is
geodesic, rotation-free and obeys the strong energy condition
$(\rho + 3p)\geq 0$, a certain generalisation of the deceleration
parameter q must be always non-negative. But even with the
perturbation considered by Kolb et al the vorticity vanishes and
consequently Kolb's claim is flawed. On the other hand, Iguchi et
al \cite{iguchi} did obtain simulated acceleration in Lemaitre –
Tolman (L–T) models with $\Lambda = 0$ that obey  the conditions
set by Hirata et al, which subsequently led Vanderveld et
al\cite{van} to draw attention to this apparent contradiction
between these two conclusions. However in a recent communication
Krasinski et al \cite{ kra} showed that L–T models that simulate
accelerated expansion also contain a weak singularity, and in this
case the derivation of HS breaks down. In addition to this, there
are other singularities that tend to arise in L–T models, and
Vanderveld et al have failed to find any singularity-free models
that agree with observations. So the apparent contradiction is
resolved. On the other hand Hansson etal \cite{hansson} argued
that when taking the real, inhomogeneous and anisotropic matter
distribution in the semi-local universe into account, there may be
no need to postulate an accelerating expansion of the universe at
all despite recent type Ia supernova data. In fact inhomogeneous
structure
formation may alleviate need for accelerating universe.\\
 Here we would like to understand if an
inhomogeneous spacetime, filled with incoherent matter, can be
turned into an accelerating universe at later times in the
framework of a higher dimensional spacetime. The inhomogeneities
considered in the present investigation may arise during an early
inflationary stage when quantum mechanical fluctuations of the
geometry and of the inflaton field are inside the Hubble radius.
Depending upon the parameters of the inflationary phase, the
initial quantum fluctuations will be amplified leading to a
quasi-flat spectrum of curvature perturbations that accounts,
through the Sachs-Wolfe effect, for the tiny temperature ripples
detected in the microwave sky by several experiments.\\\\
As pointed out earlier from Raychoudhury
 equation it can be shown that in a dust dominated universe there must
 be a non-vanishing vorticity in order to obtain a negative deceleration
 parameter. This conclusion negates Kolb's idea of presenting  inhomogeneity
  as the possible candidate to explain late time acceleration. Also the possibility
 that the full non-perturbative solutions of the Einstein's equation for
  inhomogeneous model can exhibit accelerated expansion was recently
  proved wrong by Alnes et al \cite{5} who tried to examine whether spherically
symmetric inhomogeneous universe with dust accepts negative
deceleration parameter and showed that no physically realistic solution will allow that.\\
Following Alnes et al we are motivated to see whether introduction
of extra spatial dimensions in the inhomogeneous dust distribution
can give any additional input in the direction of explaining
acceleration in the recent past. Multidimensional space-time is
believed to be particularly relevant in the context of cosmology.
Moreover in a recent communication~\cite{milton} it is argued that
quantum fluctuations in 4D spacetime do not give rise to dark
energy but rather a possible source of the dark energy is the
fluctuations in the quanutum fields including quantum gravity
inhabiting extra compactified dimensions. Here we have the extra
advantage that the exact  solution for a spherically symmetric
inhomogeneous distribution of a matter dominated universe for an
(n+2)-dimensional space-time with $n\geq2$ is earlier given by us
\cite{6}. The solution here also carries two free, spatially
dependent functions $f$ and M.\\\ The motivation of the present
work is twofold. Firstly we do away with the assumption of any
extraneous scalar field with faulty energy conditions but rather
confine ourselves to a clearly physical parameter- inhomogeneous
distribution. Secondly inspired by many recent successes of higher
dimensional theories we here take a $(n+2)$ dimensional spacetime
as our geometry. However, one should point out at the outset that
although the spacetime we have taken here for simplicity has been
utilised in the literature by a number of authors in the past(for
example, see \cite{myer} and references therein) it is open to
criticism and needs further refinement in future work. We have
organised our paper as follows: We develop the mathematical tools
in section 2 and following analogous 4D results define the
relevant astrophysical parameters suitable for the
  inhomogeneous, anisotropic model. This particularly applies
  to Hubble parameter, which unlike the homogeneous, isotropic
   case lacks a precise definition. Without exactly solving the field
   equations here we have been able to show in a general way that with
   realistic matter field the average volume expansion should always be
   decelerating. But if the radial expansion decelerates fast enough
   the angular expansion accelerates even in pure dust case and vice versa.
   But the presence of extra dimensions does not help matters, rather it
   acts as an impediment. With the help of our exact solution  we
   find exact expression of the deceleration parameter in section 3 to see that
   no acceleration is possible in this case. The above result is
   confirmed in section 4 with the help of wellknown Raychaudhuri
   equation. The paper ends with a discussion in section 5.\\

\section{Mathematical formalism}

The $(n+2)$ dimensional metric for a spherically symmetric
inhomogeneous space-time was first given by given by Banerjee et
al \cite{6}
\begin{equation}\label{a}
  ds^{2}= dt^{2}- \frac{R'^{2}(r,t) }{1 + f(r)}~dr^{2}-R^{2}(r,t)~dX_{n}^{2}
\end{equation}
where $dX_{n}^{2}$ represents an n-sphere with
\begin{equation}
dX_{n}^{2}=d\theta_{1}^{2}+\sin^{2}\theta_{1}d\theta_{2}^{2}+...\\
+\sin^{2}\theta_{1}\sin^{2}\theta_{2}...\sin^{2}\theta_{n-1}d\theta_{n}^{2}
\end{equation}
and the scale factor, $R(r,t)$  depends both on space and radial
coordinates (r,t) respectively. A prime overhead denotes
${\partial}/{\partial}r$ and a  dot  denotes  ${\partial}/{\partial}t$.\\

Here  $f(r)$ is an arbitrary function of $r$ associated with the
curvature of $t-const.$ hypersurface subject to the
restriction\begin{equation} 1+f(r)>0\end{equation} For
mathematical simplicity we here take the higher dimensional metric
with the topology $R^{n+2}$, which is however not very realistic.
As mentioned in the Introduction this type of metric, although
frequently used in literature \cite{myer}, suffers from the
disqualification that we do not get reduction of physical
quantities(say, deceleration parameter) for an effective 4D
universe as dimensional reduction is not possible. In what follows
we shall consider a dust distribution and all the dimensions
including the extra ones are treated on same footing. The original
4D metric was first studied by Lemaitre, Tolman and Bondi and
later has been used in various astronomical and cosmological
contexts. The space time (1) is a generalisation of the well known
LTB metric for the $(n+2)$ dimensions. Relevant to point out that
it reduces to the the $(n+2)$ dim. generalised FRW metric given
earlier by Chatterjee etal \cite{mnras}  in the limit $R(r,t) =
a(t)r$ and $f(r) = kr^{2}$ where $a$ is the FRW scale factor and
$k$ is the curvature constant.\\ A comoving coordinate system is
taken such that $ u^{0}=1, u^{i}= 0 (i = 1, 2, ....n+1)$ and
$g^{\mu \nu}u_{\mu}u_{\nu}= 1$ where $u_{i}$ is the
(n+2)-dimensional velocity. The energy momentum tensor for a dust
distribution in the above defined coordinates is given by
\begin{equation}
T^{\mu}_{\nu} = \rho_{M}(r,t)\delta_{0}^{\mu}\delta_{\nu}^{0} -
\rho_{\Lambda}\delta_{\nu}^{\mu}
\end{equation}
where $\rho_{M}(r,t)$ is the matter density  and we have also kept
the vacuum energy $\rho_{\Lambda}$ for generality. The fluid
consists of successive shells marked by r, whose local density ñ
is time-dependent. The function R(t, r) describes the location of
the shell marked by r at the time t. Through an appropriate
rescaling it can be chosen to satisfy the gauge
\begin{equation} R(0, r) = r\end{equation} The metric(1) with
Einstein's field equations and energy momentum tensor given by
equation (4) gives the following independent differential
equations as
\begin{equation} \frac{n(n-1)}{2}~\frac{\dot{R}^{2}- f(r)}{R^{2}}
+ \frac{n}{2}~\frac{2\dot{R'}\dot{R}- f'(r)}{RR'} = 8\pi
G(\rho_{M}+ \rho_{\Lambda})
\end{equation}
\begin{equation}
\frac{n(n-1)}{2}~\frac{\dot{R}^{2}- f(r)}{R^{2}}+
n\frac{\ddot{R}}{R}= 8\pi G \rho_{\Lambda}
\end{equation}
One can integrate the last equation by defining $\dot{R}= p(R)$
such that the scale factor $R$ itself becomes the independent
variable. The equation (7) now reduces to a Bernoulli type first
order differential equation as
\begin{equation}
p' = g(R)p + h(R)p^{-1}
\end{equation}
where $g(R) = \frac{1-n}{2}\frac{1}{R}$  and $ h(R)= \frac{8\pi G
\rho_{\Lambda}}{n}R + \frac{(n -1)f}{R}$. Following the standard
method of solving this type of equation we finally get
\begin{equation}
\frac{\dot{R}^{2}}{R^{2}} = \frac{M(r)}{R^{n +1}}+
\frac{f(r)}{R^{2}} + \frac{16\pi G\rho_{\Lambda}}{n(n+1)}
\end{equation} where $M(r)$ is an arbitrary function of
integration and depends on $r$. From the above equations it
further follows that for pure matter field ($\rho _{\Lambda} = 0)$
the equation (9) reduces to
\begin{equation}
\dot{R}^{2}= f(r)+ \frac{M(r)}{R^{n-1}}
\end{equation}
which , again, gives
\begin{equation}
\frac{nM'}{2R'R^{n}}= 8\pi G\rho_{M}
\end{equation}
such that  $M(R)$ is non negative, being a measure of the mass
content for the n-sphere upto the comoving radius r. The
generalized mass function M(r) of the fluid can be chosen
arbitrarily. It incorporates the contributions of all shells up to
r and determines the energy density through equation (11). Because
of energy conservation M(r) is independent of t. Moreover the
actual dependence of the arbitrary functions $M(r)$ and $f(r)$ are
determined by the specific nature of the inhomogeneities of our
model considered.\\ Inhomogeneous distributions being always a bit
obscure one can attempt greater transparency via usage of familiar
physical quantities like Hubble constant, H and also the density
parameter, $\Omega_{M}$ from equation (9) through analogy with the
wellknown homogeneous FRW equations generalised to $(n+2)$
dimensions
\begin{equation}
H^{2}(t) = \frac{\dot{a}^{2}}{a^{2}} = H_{0}^{2}
~[\Omega_{M}(\frac{a_{0}}{a})^{n +1}+ \Omega_{\Lambda}+ (1-
\Omega_{M}- \Omega_{\Lambda})(\frac{a_{0}}{a})^{2}]
\end{equation}
where $a_{0}= a(t_{0})$ is the current value of the scale factor.
Comparing equations(9) and(12) one can now identify the local
Hubble constant as
\begin{equation}
H(r,t)=\frac{\dot{R}(r,t)}{R(r,t)}
\end{equation}
and local matter density via
\begin{equation}
M(r)= H_{0}^{2}~\Omega_{M}(r)R_{0}^{2}(r)
\end{equation}\\
where
\begin{equation}
f(r)= H_{0}^{2}(r)~[\Omega_{M}(r) + \Omega_{\Lambda}(r)-
1]R_{0}^{2}(r)
\end{equation}
where $R_{0}= R(r,t_{0}), H_{0}= H(r,t_{0})$~ and
$\Omega_{\Lambda}(r)) = \frac{16\pi
G\rho_{\Lambda}}{n(n+1)H_{0}^{2}(r)}$~ The equation (9) can now be
recast as
\begin{equation}
H^{2}(r,t) = H_{0}^{2}(r)~[ \Omega_{M}(r)~
(\frac{R_{0}}{R})^{n+1}+ \Omega_{\Lambda}(r)+ \Omega_{c}(r)~
(\frac{R_{0}}{R})^{2}]
\end{equation}
where $ \Omega_{c}(r) = 1 - \Omega_{_{\Lambda}}(r)-\Omega_{M}(r)$.
One should note at this stage that although seemingly same the
essential difference of this generalised LTB expressions from the
standard FRW case is that all the quantities here depend on
spatial coordinate also and when $n=2$ they reduce to the 4D LTB
case. This is true even for the gauge freedom exercised in (5).
While for the FRW case the present value of the scale factor
$a(t_{0})= a_{0}$ can be chosen to be any positive number the
analogous LTB scale factor $A(r~ t_{0})$ can be chosen to be any
smooth and invertible positive function. From the relations (6)
and (7) a little mathematical exercise yields
\begin{equation}
\frac{n}{n+1}\frac{\ddot{R}}{R}+ \frac{1}{n+1}\frac{\ddot{R'}}{R'}
= -\frac{8\pi G}{n(n+1)}[(n-1)\rho_{M}- 2\rho_{\Lambda}]
\end{equation}\\
The equation(17) tells us that the average acceleration in our
case is generally negative unless $\rho_{\Lambda}>\frac{n
-1}{2}\rho_{M}$.  Given the fact that the current observational
value of $\Lambda$ is too small the last inequality is a remote
possibility. It has not also escaped our notice that even for a
pure matter dominated model the radial acceleration
$(\frac{\ddot{R'}}{R})$ is possible if our angular scale factor is
decelerating fast enough and vice versa. Another important fallout
is the role of the extra dimensions in the dynamical process. The
equation(16) shows that as the no. of dimensions increases  the
possibility of achieving acceleration in the model further
recedes. We shall subsequently see that this result also follows
from the well known RayChowdhuri equation as well.

The volume expansion rate for our $(n+2)$ dim. metric is  defined
through the (n+2)-velocity of the fluid, $u^{a}$ as
\begin{equation} (n +1)H = u^{a}_{; a}=u_{a ; b}~ g^{ab}=u_{a ; b}~ h^{ab}
\end{equation}
where
\begin{equation}
h^{ab}= g^{ab}+ u^{a}u^{b}
\end{equation}

While the definition works perfectly well for a FRW like
homogeneous distribution of matter it is always a bit ambiguous to
define the deceleration parameter of an inhomogeneous anisotropic
model because the relation (18) does not take into account the
directional preference of the matric. For example, Tolman-Bondi
has a preferred direction, being the radial one. We can still give
an operational definition to the average \emph{volume
acceleration} of our model. For inhomogeneous model the
directional preference need to be emphasized in the expression for
expansion.  We define a projection tensor $t^{ab}$ that projects
every quantity perpendicularly to the preferred spacelike
direction $s^{a}$ (and of course the timelike vector field,
$u^{a}$) such that
\begin{equation}
t^{ab} = g^{ab}+ u^{a}u^{b} - s^{a}s^{b}= h^{ab}- s^{a} s^{b}
\end{equation}
For our metric (1), \begin{equation} p^{a}= \frac{\sqrt{1+
f(r)}}{R'}\bigtriangledown\end{equation} and the tensor projects
every physical quantity in a direction $\bot$ to $s^{a}$. One can
now define the invariant expansion rates as
\begin{equation}
H_{r}= u_{a ; b} s^{a}s^{b} = \frac{\dot{R'}}{R'}
\end{equation}

\begin{equation}
H_{\bot}= \frac{1}{n}u_{a ; b}t^{ab}= \frac{\dot{R}}{R}
\end{equation}
so that
\begin{equation}
H = \frac{n}{n+1}H_{\bot} + \frac{1}{n+1}H_{r}
\end{equation}
Evidently the above definition gives a sort of averaging over the
various directions for our anisotropic model.
\\
If one relaxes the condition of any particular preferred direction
(like the radial one as in LTB model) one can explore the
definition of the Hubble parameter in a more transparent way
considering its directional
dependence \cite{partovi} as follows:\\
\begin{equation}
H = \frac{1}{n+1}u^{a}_{_{a}} + \sigma_{a b}J^{a}J^{^{b}}
\end{equation}
where $\sigma_{a b}$ is the shear tensor and $J^{^{a}}$ a unit
vector pointing in the direction of observation. For an observer
located  away from the centre of the configuration it gives for
our LTB case
\begin{equation}
H = \frac{\dot{R}}{R} + \left(\frac{\dot{R'}}{R} -
\frac{\dot{R}}{R}\right) cos^{2}\theta
\end{equation}
 where $\theta$ is the angle between the radial
direction through the observer and the direction of observation.
Naturally when the two directions coincide, $\theta=0$ we get
$H=H_{_{r}}$ and for $\theta=\frac{\pi}{2}$ it is $H =
H_{\theta}$. A definition of deceleration parameter in a preferred
direction can also be given in terms of the expansion of the
Luminosity distance $D_{L}$ in powers of redshift of the incoming
photons. For small $z$ one gets
\begin{equation}
q = - \dot{H}\frac{d^{2}D_{L}}{dz^{2}}+ 1
\end{equation} For $\theta =0$ and $\theta=\pi/2$ the acceleration is
respectively
\begin{equation}
q_{r} = -
\left(\frac{R}{\dot{R'}}\right)^{2}\left[\frac{\ddot{R'}}{R}-\frac{\sqrt{1
+ f}} {\dot{R'}}\left(\frac{\dot{R'}}{R}\right)'\right]
\end{equation}

\begin{equation}
q_{\bot}= - \left(\frac{R}{\dot{R}}\right)^{2}\frac{\ddot{R}}{R}
\end{equation}
We shall subsequently see in section 4 that deceleration parameter
defined this way has an important difference from what we later
get in equation (48). Here the parameters do not depend solely on
local quantities as opposed to the acceleration parameter of (48).
For example we get via  equation(10)
\begin{equation}
q_{\bot}= (n-1)\frac{M(r)}{R^{n+1}}\frac{1}{H_{\bot}^{2}}
\end{equation}
Thus the equation (30) tells us that here the deceleration
parameter $q_{\bot}$ depends on the total mass function and not on
the local energy density of (48).

All the expressions reduce to the familiar Tolman-Bondi case when
 $n=2$. It is very difficult to find a general solution of the
equation(10). But for the 4D case the scale factor may be
expressed in a parametric form for f(r)~(-1. 0, +1). But for $n>2$
the equation can not be expressed in a parametric form as the
system of equations become elliptic.
 \\\\
 \section{Higher dimensional LTB metric}
\textbf{Case I (f~=~0)}\\
 Since the WMAP data\cite{13}shows that the universe is spatially flat to
 within a few percent we can take $f=0$ to get the globally flat solution in
 $(n+2)$ dimensions as \\
\begin{equation}
R=\left(\frac{n+1}{2}\right)^{\frac{2}{(n+1)}}M^{\frac{1}{(n+1)}}
(t-t_{0})^{\frac{2}{(n+1)}}
\end{equation}
where $t_{0}(r)$ is some arbitrary integration function of r.\\
Hence
 \begin{eqnarray}
 \dot{R}&=&\left(\frac{n+1}{2}\right)^{\frac{(1-n)}{(n+1)}}M^{\frac{1}{(n+1)}}
 (t-t_{0})^{\frac{1-n}{(n+1)}}\\
 \ddot{R}&=&\left(\frac{n+1}{2}\right)^{\frac{(1-n)}{(n+1)}}M^{\frac{1}{(n+1)}}
 \left(\frac{1-n}{1+n}\right)(t-t_{0})^{-\frac{2n}{(n+1)}}
\end{eqnarray}
So,
\begin{equation}
 q = -\frac{\ddot{R}}{R}\frac{1}{(\frac{\dot{R}}{R})^{2}}=(n-1)\left(\frac{n+1}{2}\right)^{\frac{n}{n+1}}
\end{equation}
Thus $n<1$ is needed for acceleration. Further the fig 1 shows
that with dimensions deceleration also increases. So addition of
extra dimensions is, at least for this model, counterproductive.
 \\
\begin{figure}
    \begin{center}
    \includegraphics[width=6cm]{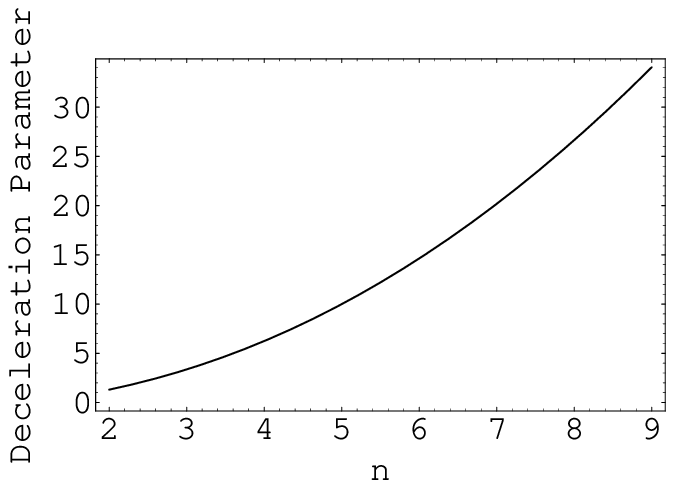}
    \includegraphics[width=6cm]{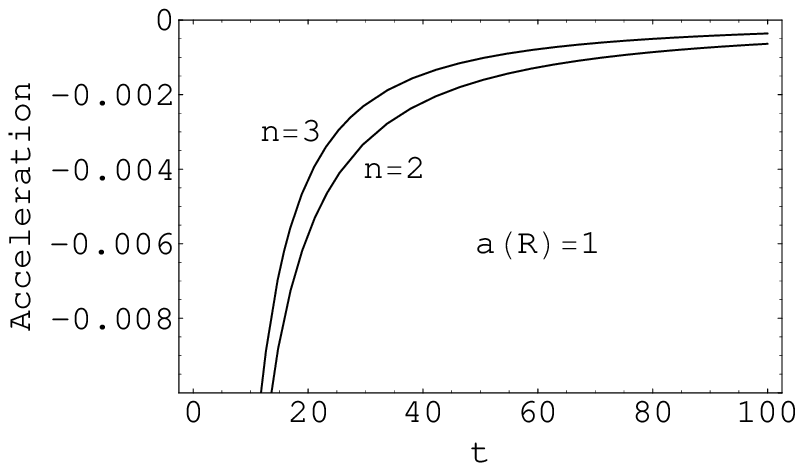}
    \caption{$q$ vs $n$ and $\ddot{R}$ vs $t$}\label{qvst}
    \end{center}
\end{figure}\\\textbf{Case II}$ (f \neq 0)$ \\
As is well known that in the 4D spacetime we do not get the
solution of the equation(10) in a closed form for $f(R)\neq 0$. At
best one gets a parametric form of solutions. But a positive thing
in higher dimensional cosmology lies in the fact that at least in
5D case($n=3$)we get an analytic solution as\\
\begin{equation}
 R= \left[f(t-t_{0})^{2}-\frac{M}{f}\right]^{\frac{1}{2}}
\end{equation}
The above equation has been utilised to extensively study the
shell crossing and shell focussing singularity generally
associated with any inhomogeneous collapse(see ref \cite{6} for
thorough discussion) such that the 5D case $(n=3)$gives
\begin{equation}
 \ddot{R}
 = -M\left[f(t-t_{0})^{2}-\frac{M}{f}\right]^{-\frac{3}{2}}\end{equation}\\
 \begin{equation}
 q = -\frac{\ddot{R}}{R}\frac{1}{(\frac{\dot{R}}{R})^{2}}=\frac{M}{{f}^{2}(t-t_{0})^{2}}
\end{equation}
Thus for deceleration parameter to be negative, M, which is
identified as the mass density parameter must be negative. But
equation $(11)$ demands that $M'(r)$ must be positive which in
turn demands $M(r)\geq0$. Hence dust dominated spherically
symmetric model even in higher dimension does not
 allow acceleration with a physically realistic matter field.\\\
 \section{Generalised Raychaudhuri Equation}

It may not be out of place to address the situation discussed in
the last section with the help of the well known Ray Chaudhuri
equation\cite{ray}, which in general holds for any cosmological
solution based on Einstein's gravitational field equations. In an
earlier work \cite{sujit} one of us extended the Ray Chaudhuri
equation, the null congruence condition and also the focussing
theorem to (n +2) dimensions to study how inclusion of extra
spatial dimensions alters the possibility of occurrence of
singularities in physically realistic situations. We later applied
the generalised equation to specific higher dimensional
cosmological problems. Before writing the generalised equation
proper the following definitions in (n+2) dimensions may be in
order
\begin{equation}
\sigma_{\mu\nu} = u_{(\mu ;~ \nu)}- \frac{1}{n+1}\Theta(g_{\mu
\nu}
 + u_{(\mu}u_{\nu )}) +   u_{(\mu}u_{\nu); j}u^{j}
\end{equation}
where for our model given by equation(1)
\begin{equation}
\Theta = u^{\mu}_{; ~\mu}= \frac{\dot{R'}}{R'}+ n\frac{\dot{R}}{R}
\end{equation}
\begin{equation}
\omega_{\mu\nu}= u_{~[\mu;\nu]} - a_{[\mu}u_{\nu]}
\end{equation}
Here $\Theta$ is the rate of volume expansion and $\sigma_{\mu
\nu}$ is the rate of shearing.\\ For our LTB metric generalised to
$(n+2)$ dimensions this gives

\begin{equation}
\sigma^{1}_{1} = \frac{n}{n+1}\left(\frac{\dot{R'}}{R'}-
\frac{\dot{R}}{R}\right)
\end{equation} and
\begin{equation}
\sigma^{2}_{2}= \sigma^{3}_{3}=...........=\sigma^{n+1}_{n+1}=
-\frac{1}{2}~ \sigma^{1}_{1}
\end{equation}
This finally gives for our $(n+2)$ dim. model the shear scalar as

\begin{equation}
\sigma^{2} = \sigma_{\mu\nu}\sigma^{\mu \nu} = \frac{n^{2}(n+4)}
{4~(n+1)^{2}}\left(\frac{\dot{R'}}{R'}-
\frac{\dot{R}}{R}\right)^{2}
\end{equation}
This reduces to the 4D value \cite{enquist} of shear scalar for
$n=2$.
 It may be noted that here $(\mu, \nu)$ run from zero
to (n+1).\\With the help of the above definitions we get after a
long but straight calculation the well known Ray Chaudhuri
equation generalised to (n+2) dimensions as
\begin{equation}
\Theta_{,\mu}v^{\mu} = \dot{v}^{\mu}_{; \mu} - 2(\sigma^{2}-
\omega^{2})-\frac{1}{n +1}~~ \Theta^{2} + R_{\nu
\alpha}v^{\nu}v^{\alpha}
\end{equation}
In view of Einstein's equations the last term in the equation(44)
may be replaced by $-8\pi G[ T_{\nu \alpha} - \frac{1}{n}T]$,
where G is now the (n+2) dim. gravitation constant. With matter
field expressed in terms of mass density and 3D and higher
dimensional pressures the (n + 2) dimensional  Ray Chaudhury
equation is finally  given by,
\begin{equation}\label{a}
  \dot{\Theta}=-2(\sigma^{2}-\omega^{2})-\frac{1}{(1+n)}\Theta^{2}-\frac{8\pi G}
  {n}[(n- 1 )\rho+3p+(n-2)p_{e}]
\end{equation}
in a  co moving reference frame. Here $p$ and $p_{e}$ are the 3D
and extra dimensional isotropic pressure respectively.\\
With the help of equation(24) we get an expression for an
effective deceleration parameter as
\begin{equation}\label{b}
 q= -\frac{\dot{H} + H^{2}}{H^{2}} = -1-(n +1)\frac{\dot{\Theta}}{\Theta^{2}}
\end{equation}
 which allows us to write,
\begin{equation}\label{c}
  \Theta^{2}q = 2(n + 1)\sigma^{2}+\frac{8\pi G}{n}(n+1)[(n-1)\rho
  + 3p+(n-2)p_{d}]
\end{equation}
One should look at this relation with caution. While working in
higher dimensional cosmology it is envisaged that following
dimensional reduction as the 3D expands the extra dimensions
shrink to a microscopic size as to be invisible with the existing
low energy limit so that all physical quantities (e.g.
deceleration parameter) become effectively four dimensional. As we
have to sacrifice dimensional reduction for mathematical
simplicity nothing of that sort takes place in our case, which is
definitely a major defect of the present analysis and need to be
rectified in a future work.
 Although in reference \cite{kolb} Kolb et al claimed that the vorticity
 of the LTB metric is nonvanishing our careful calculations show that $\omega =0$
even in the
  generalised LTB model. As we are dealing with a LTB cosmology  the fluid
  is vorticity free and
 pressure less in standard $4D$ as well as higher dimension. So we finally
 get for the particular case of generalised LTB model(p = $p_{e}$ = 0)
\begin{equation}\label{c}
\Theta^{2}q = 2(n + 1)\sigma^{2}+\frac{8\pi G}{n}(n+1)(n-1)\rho
\end{equation}
As we can see $\rho$, $\Theta^{2}$ and $\sigma^{2}$ all are
positive
 we can conclude $q$ is positive. So the addition of extra dimensions
 has no qualitative impact in determining the signature of the deceleration
 parameter. Let us analyse the last equation a little more
 thoroughly. For the flat$(n +1)D$ space with $f=0$ a long but straightforward
calculation shows that
\begin{equation}
 \theta=\frac{1}{(t-t_{0})}\left[\frac{2(\frac{M'}{M})(t-t_{0})-t_{0}'}
 {(\frac{M'}{M})(t-t_{0})- t_{0}' }\right]
\end{equation}

and
\begin{equation}
 \sigma^{2}=\frac{n^{2}(n+4)}{4 (n+1)^{2}}(t-t_{0})^{-2}\left[\frac{t'_{0}}
 {(t - t_{0})\frac{M'}{M}- t'_{0}}\right]^{2}
\end{equation}

\begin{figure}
    \begin{center}
    \includegraphics[width=6cm]{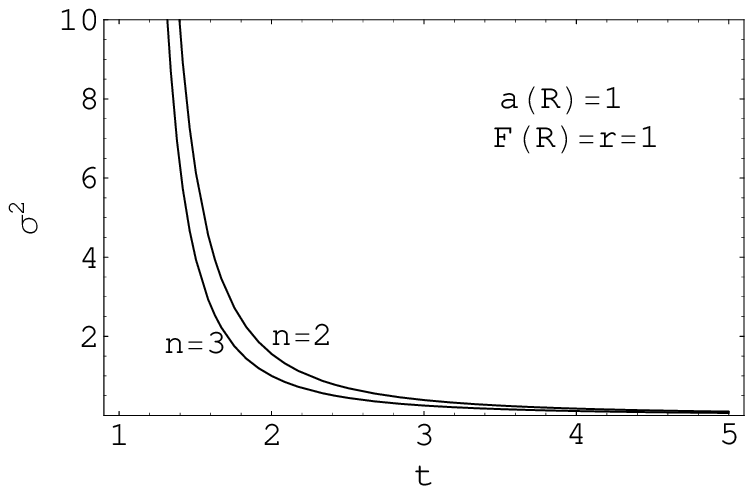}
    \includegraphics[width=6cm]{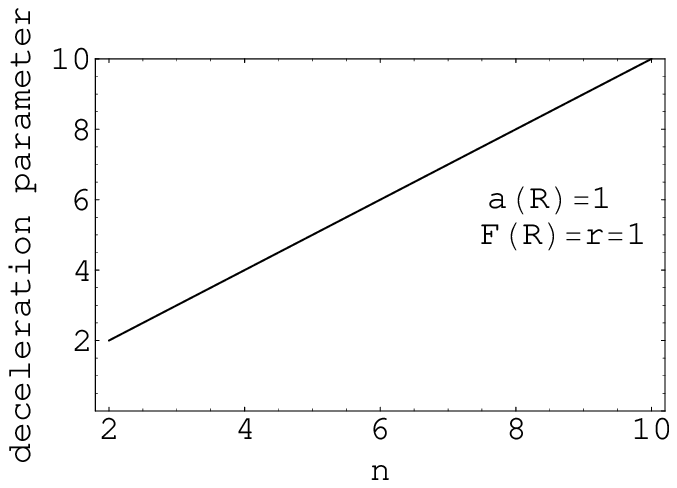}
    \caption{Shear scalar vs. t and $q$ vs $n$}\label{avst}
\end{center}
\end{figure}

Thus with zero curvature, the expansion scalar does not depend on
the dimension but falls off as $1/t$ with time. This, in our
opinion, is not a generic result but holds for this specific case
only. But the shear scalar does depend on the dimension, falling
off faster as dimension increases. From equation(50) it appears
that shear vanishes when $t'_{0}=0$. In fact in the standard
treatment of the LTB metric the radial coordinate is taken such
that the initial energy density is perceived as homogeneous.
Moreover $f(r)$ is taken as zero to provide consistency with the
matter dominated flat model. The inhomogeneity is introduced
through the function $t_{0}(r)$ that appears in the integration of
the field equations and determine the local bigbang time. When we
plug in all these expressions in the equation (48) we find that
the deceleration parameter still increases with the increase in
dimension as evident from the fig.2 and we can not get any flip in
its signature. This finding is also in line with our observation
in the last section.\\
Our investigations in both the sections leave the impression that
physically acceptable inhomogeneous models with a realistic matter
field are unable to account for observations. But we argue that
inhomogeneous models like LTB include FRLW models as a subclass.
Thus, if the FLRW models are considered good enough for cosmology,
then the L–T models can only be better: they constitute an exact
perturbation of the FLRW background, and can reproduce the latter
as a limit with an arbitrary precision. The most serious
misconception emanates from the realm of accelerating universe
itself. One can have a very good fit with observations even with
$q <0$\cite{x,y}. An important point to remember is that the
expansion rate of the universe is not a quantity that is directly
observable. It is inferred indirectly through the observations
only after one assumes a model for the expansion of the universe.
Thus, instead of trying to explain why the expansion is
'accelerating`, one should try to explain the data themselves
directly in terms of observable quantities. One such observable is
the Luminosity distance- redshift relation, which is probed by
supernova observations. While within the framework of homogeneous
and isotropic model this relation can only be explained if the
expansion rate is accelerating, this is not the case with
inhomogeneous models like the LTB one. As discussed in section x
the added freedom of having a position dependent expansion in LTB
models allows one to explain the data without the need for the
expansion to accelerate locally. The explanation will then be that
the expansion rate is highest at $r=0$ and decreases with distance
from the centre, since the oldest supernova are also furthest away
\cite{cel}. In fact Iguchi et al \cite{iguchi} show that
 models with $M(r)>0$ can reproduce the relationship of a $\Lambda CDM$
 model upto $z\sim 1$, but not for higher redshifts. In this work,
 however, we have not so far tried to confront our higher dimensional
 model with actual SNIa observations and leave that for a future work to
 ascertain if the addition of extra dimensions makes a better fit with
 experimental data. \\
\section{Discussion}
The main shortcoming of the present analysis is the choice of the
topology of the spacetime itself. Unlike the brane inspired models
where observable matter is trapped on a brane but the extra space
has macroscopic size or the `\emph{space-time-matter}' theory of
Wesson \cite{wess} where  extra dimensions is a product space with
noncompact macroscopic size but matter field is a manifestation of
the higher dimensional effect  we here, for simplicity,
\emph{naively} take a spacetime with topology $R^{n+2}$ such that
all the spatial dimensions are taken on the same footing. This may
be the case in the very early universe before the cosmology
underwent the compactification transition. But here we are
discussing a late scenario where the universe is manifestly 4D
with extra dimensions presumably compactified below planckian
length. As our spcetime is not amenable to the desired feature of
dimensional reduction, in a certain sense the relevance of this
model for the current scenario is somewhat obscure. As a future
exercise one should address the problem such that the extra space
forms a compact manifold with symmetry group G such that (n+1)
spatial symmetry group is a direct product of $O(3)\times G$ and
not the simple $O(n+1)$.

Aside from the above defect and presumably a host others we have
made a preliminary  attempt to see if the presence of
inhomogeneities or extra spatial dimensions separately or jointly
can achieve late acceleration without the aid of any extraneous
unphysical quintessential scalar field or an evolving cosmological
constant. In fact the notion of accelerating expansion has
subtleties in inhomogeneous cosmology. The conventional definition
of the deceleration parameter through equation (17) is bizarre, if
not confusing. It actually accounts for local volume increase
during the expansion. One should note that for inhomogeneous or
anisotropic model it pertains to a sort of averaging over various
directions. For LTB model we find that this average expansion rate
is always decelerating for positive energy density. This is also
corroborated by the wellknown Raychaudhuri equation. It has not
escaped our notice that presence of extra dimensions actually
favours the decelerating process. However, without explicitly
solving the field equations we find that the radial or the angular
acceleration is possible even in pure dust distribution if any one
of them decelerates fast enough even in LTB model. In recent past
several authors \cite{darabi} for example, Darabi as also
Szydlowski et al have claimed to get accelerating model with the
help of extra dimensions. While the first case deals with
primordial inflation, the second author has assumed a negative
pressure in the fifth dimension $p_{5}< 0$ in his 5D metric to
achieve the late acceleration. This  is already discussed by
Panigrahi and Chatterjee  in an earlier work \cite{dp}. But
introduction of an unphysical negative pressure even in extra
dimension does not lead us any nearer to real physics. It only
shifts the unphysical input from the 3D space to the extra
internal fifth dimension. To end the section a final remark may be
reemphasised regarding the apparent accelerated expansion of the
universe. To explain the SNIa observations the concept of
accelerated expansion of the universe need to be invoked only for
a FRW type of model. But one should point out that the Luminosity
distance- Redshift relation, not the accelerated expansion is the
quantity that can be  directly measured. And within inhomogeneous
models, as discussed at the end of the last section  one gets
better fit without the need to introduce the local accelerated
expansion and consequent hypothesis of any extraneous, unphysical
matter field with large negative pressure.\\In the coming years we
all hope to learn much more about inflation and observed cosmic
acceleration of the universe (attributed to dark energy) from the
highly refined computations and sophisticated observations. More
data from different cosmic probes as also LHC experiments at CERN
might help to address questions about the early universe and the
high energy frontiers. The years ahead will certainly bring even
more twists, breakthroughs and surprises in gravity and cosmology
research and give more insight into the vexed problem of origin of
dark energy.\\
\textbf{Acknowledgments} \\The financial support of UGC, New Delhi
 in the form of a MRP award as also a Twas Associateship award, Trieste
 is  acknowledged. I also thank Prof. P. Letelier, IMPEEC, Campinas University,
 Brasil for local hospitality where part of the work is done. The helpful
 comments from the anonymous referee, leading to improvements of the
 original version is also acknowledged.\\

\end{document}